\documentclass[twocolumn,prb]{revtex4}

\usepackage{picinpar}
\usepackage{array}
\usepackage{subfigure}
\usepackage{amsmath}
\usepackage{multirow}
\usepackage[dvips]{graphicx}
\usepackage{amssymb}

\begin{document}

\title{Resilience of gas-phase anharmonicity in the vibrational response of adsorbed carbon monoxide and breakdown under electrical conditions}

\author{Ismaila Dabo}
\email{daboi@cermics.enpc.fr}
\affiliation{Universit\'e Paris-Est, CERMICS, Project-team INRIA Micmac, 
6 \& 8 avenue Blaise Pascal, 77455 Marne-la-Vall\'ee, France}

\begin{abstract}
In surface catalysis, the adsorption of carbon monoxide on transition-metal electrodes represents the prototype of strong chemisorption. Notwithstanding significant changes in the molecular orbitals of adsorbed CO, spectroscopic experiments highlight a close correlation between the adsorbate stretching frequency and equilibrium bond length for a wide range of adsorption geometries and substrate compositions. In this work, we study the origins of this correlation, commonly known as Badger's rule, by deconvoluting and examining contributions from the adsorption environment to the intramolecular potential using first-principles calculations. Noting that intramolecular anharmonicity is preserved upon CO chemisorption, we show that Badger's rule for adsorbed CO can be expressed solely in terms of the tabulated Herzberg spectroscopic constants of isolated CO. Moreover, although it had been previously established using finite-cluster models that Badger's rule is not affected by electrical conditions, we find here that Badger's rule breaks down when the electrified surface is represented as a periodic slab. Examining this breakdown in terms of anharmonic contributions from the effective surface charge reveals limitations of conventional finite-cluster models in describing electrical conditions at metal electrodes.  
\end{abstract}

\maketitle

\section{Introduction}

The significance of carbon monoxide adsorption at transition-metal surfaces extends beyond its central relevance to catalytic exhaust control and electrochemical energy conversion. Carbon monoxide is a cornerstone for fundamental research in surface science and electrochemistry\cite{Somorjai1996,  IwasitaNart1997, Rupprechter2007} that has been sufficiently characterized spectroscopically to serve now as a reliable reference in measuring the potential of zero charge of metal electrodes, \cite{Weaver1998, GomezCliment2000, ClimentAttard2002, FriedrichDaum2003, Cuesta2004, ClimentGarcia-Araez2006} identifying coadsorbates in complex surface environments, \cite{ClavilierAlbalat1992, ClimentGarcia-Araez2006, SubbaramanStrmcnik2010} elucidating adsorption phenomena, \cite{ClavilierAlbalat1993, SubbaramanStrmcnik2010} and probing electrical conditions in the electrochemical double layer. \cite{ClavilierAlbalat1992, ClavilierAlbalat1993, FeliuOrts1994, FriedrichDaum2003, SubbaramanStrmcnik2010} Among successful spectroscopic techniques, infrared experiments have delivered quantitative insight into the properties of CO adsorbed at catalytic electrodes, unveiling distinctive vibrational fingerprints as a function of the adsorption environment.  \cite{IwasitaNart1997, Park2002, ParkTong2002, ParkWieckowski2003, LuWhite2004, LuLagutchev2005, MaillardLu2005, LagutchevLu2006} In fact, the intramolecular frequency $\omega_{\rm e}$ of CO at close-packed metal surfaces undergoes redshifts of several hundreds of cm$^{-1}$ with increasing adsorption coordination while augmenting gradually with the nobleness of the metal substrate. \cite{GajdosEichler2004} In particular, this infrared trend has enabled for the vibrational recognition of CO adsorption sites on functional catalytic alloys. \cite{DaboWieckowski2007}

Following the seminal work of Blyholder that has set forth the theory of electron donation (the depletion of the 5$\sigma$ bonding frontier molecular orbital) and electron backdonation (the filling of the 2$\pi^*_x$ and 2$\pi^*_y$ antibonding orbitals) in discussing the electronic origins of CO chemisorption, \cite{Blyholder1964} an important corpus of theoretical and experimental literature has developed on the elucidation of C--O vibrational redshifts as a function of adsorption conditions. One salient, albeit unexpected spectroscopic feature that has attracted attention is the close correlation between the C--O equilibrium stretching frequency $\omega_{\rm e}$ and equilibrium bond length $r_{\rm e}$ upon modifying the adsorption environment. \cite{OgletreeVan-Hove1986, WasileskiWeaver2002} In quantitative terms, the linear-regression correlation ${\cal B} = \partial \omega_{\rm e}/\partial r_{\rm e}$ (the Badger slope) is negative with an absolute value in the range 6500-7500 cm$^{-1}\cdot$\AA$^{-1}$ regardless of the substrate composition and adsorption site. In this work, we examine the electronic origins of this notable spectroscopic trend and carry the analysis further to probe the predictive accuracy of widely used first-principles computational approaches in describing molecular adsorbates under electrical conditions. 

The study is organized as follows. First, we present structural and vibrational calculations for CO in the gas phase and focus on describing spectroscopic correlations for CO adsorbed on close-packed (111) metal surfaces from local and semilocal density-functional theories. Second, we introduce a quantitative intramolecular analysis to resolve anharmonic contributions from the surface environment. On the basis of this analysis, we demonstrate that the correlation between $\omega_{\rm e}$ and $r_{\rm e}$ for CO adsorbed at metal surfaces can be expressed in terms of the spectroscopic constants of gas-phase CO. Finally, we explore implications of these results in simulating local electrical conditions and interpreting adsorption phenomena at electrified metal surfaces.

\section{First-principles spectroscopy}

\subsection{Carbon monoxide in the gas phase}

\begin{table}
\caption{Spectroscopic constants of CO compared with experiment (Ref.~\onlinecite{CRC2011}).
\label{GasPhaseSpectroscopy}}
\begin{tabular*}{\columnwidth}{@{\extracolsep{\fill}}lccccc}
\hline \hline
                                            & HF        & LDA      & PBE      & Expt. \\
\hline
\\
$r_{\rm e}$ (\AA)                           & 1.087     & 1.124    & 1.141    & 1.128   \\
$\omega_{\rm e}$ (cm$^{-1}$)                & 2434.14   & 2128.59  & 2103.93  & 2169.81 \\
$\omega_{\rm e}x_{\rm e}$ (cm$^{-1}$)       & 11.47     & 12.02    & 11.91    & 13.29   \\
$B_{\rm e}$ (cm$^{-1}$)                     & 2.08115   & 1.94558  & 1.88862  & 1.93128 \\
$\alpha_{\rm e}$ (cm$^{-1}$)                & 0.01597   & 0.01774  & 0.01778  & 0.01750 \\
$D_{\rm e}$ ($10^{-6}$cm$^{-1}$)            & 6.0852    & 6.5016   & 6.08799  & 6.1216  \\
\\
\hline
\hline
\end{tabular*}
\end{table}

\begin{figure}[ht!]
\includegraphics[width=8cm]{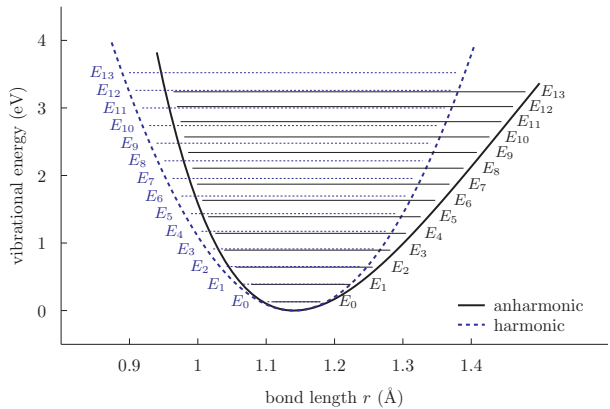}
\caption{
Comparison of intramolecular vibrational harmonic energy levels $E_\nu = hc \omega_{\rm e} (\nu+1/2)$ and anharmonic energy levels 
$E_\nu = hc\sum_{k=0}^\infty Y_{k0} (\nu+1/2)^k$ for CO in the gas phase.
\label{EnergyLevels}}
\end{figure}

\begin{figure}[ht!]
\includegraphics[width=8cm]{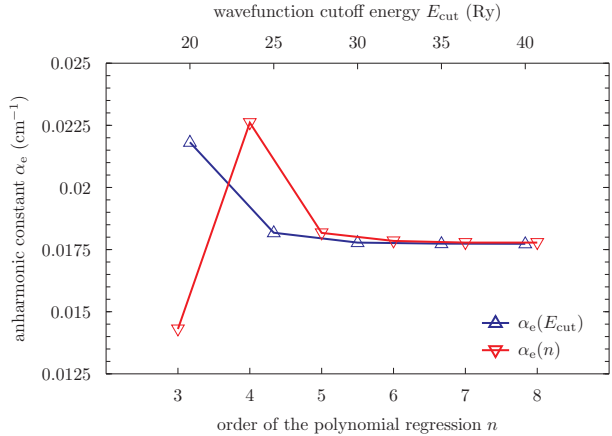}
\caption{Convergence of the Herzberg anharmonic spectrocopic constant $\alpha_{\rm e}$ for CO in the gas phase as a function
of the order of the polynomial regression of the potential energy curve (lower scale) and cutoff kinetic energy of the wave-function expansion (upper scale).
\label{HerzbergConvergence}}
\end{figure}

As a prelude to the intramolecular analysis presented in Sec. \ref{Analysis}, we report electronic-structure predictions for the structural, vibrational, and nonlinear spectroscopic properties of CO in the gas phase. For the purpose of this analysis, particular attention is paid to the determination of high-order spectroscopic constants; those account for anharmonic contributions to the Born-Oppenheimer potential energy $V(r)$ (here, the coordinate $r$ stands for the intramolecular bond length).

In Fig.~\ref{EnergyLevels}, we compare C--O vibrational energy levels evaluated within the harmonic approximation with the vibrational spectrum calculated taking into account anharmonic contributions to the potential. Within the harmonic approximation, the potential energy can be expressed in terms of the diatomic effective mass $m$, the equilibrium bond length $r_{\rm e}$, and the equilibrium stretching frequency $\omega_{\rm e}$ (in photon wavenumbers) as a quadratic function of the form
\begin{equation}
V(r)=\frac 12 mc^2(2\pi \omega_{\rm e})^2(r-r_{\rm e})^2+ \cdots .
\label{DEP}
\end{equation}
The diagonalization of the harmonic Hamiltonian yields vibrational energy levels that are separated by a constant energy gap, i.e.,
\begin{equation}
E_\nu=hc\omega_{\rm e}(\nu+1/2),
\end{equation}
where $\nu \ge 0$ denotes the vibrational quantum number.

Considering now anharmonicity, the intramolecular potential can be written in the conventional adimensional Dunham form, \cite{Dunham1932}
\begin{equation}
V(\xi)=hca_0\xi^2(1+a_1\xi+a_2\xi^2+\cdots),
\end{equation}
where $\xi=(r-r_{\rm e})/r_{\rm e}$ denotes the relative elongation of the bond and the terms $a_i$ are the coefficients of the Dunham expansion. Following the analytical treatment of Ref.~\onlinecite{Dunham1932}, anharmonic energy levels can be evaluated as
\begin{equation}
E_\nu=hc\sum_{k=0}^{+\infty} Y_{k0}(\nu+1/2)^k,
\end{equation}
where the $Y_{k0}$'s are explicitly defined as a function of the Dunham $a_i$'s. The first coefficient $Y_{00}$ corresponds to the anharmonic energy shift while the coefficient $Y_{10}$ can be identified straightforwardly to be
\begin{equation}
Y_{10}=\omega_{\rm e}.
\end{equation}
The coefficient $Y_{20}$ of the third term is typically negative and orders of magnitude smaller than $Y_{10}$; it reflects the gradual narrowing of level separations in the upper part of the vibrational spectrum. Higher-order coefficients $Y_{n0}$ are of alternating sign and decreasing magnitude. Quantitatively, anharmonicity represents a negligible energy correction at the bottom of the vibrational spectrum, affecting the vibrational zero-point energy by less than a meV for gas-phase CO. In contrast, anharmonicity becomes significant at higher vibrational energies. In fact, for $\nu \ge 10$, anharmonic corrections are on the order of one fourth of an eV, that is, comparable to the separation between energy levels.

Alternatively to the Dunham adimensional representation, in the Herzberg vibrational description, \cite{WilsonBernath2003} anharmonic levels are written in terms of the spectroscopic constants $x_{\rm e}$ and $y_{\rm e}$ as
\begin{eqnarray}
E_\nu&=&hc\Big[\omega_{\rm e}(\nu+\frac 12) - \omega_{\rm e}x_{\rm e}(\nu+\frac 12)^2 \nonumber \\
&&+ \omega_{\rm e}y_{\rm e}(\nu+\frac 12)^3 + \cdots \Big],
\end{eqnarray}   
which allows us to identify Dunham coefficients as
\begin{eqnarray}
Y_{10} & = & - \omega_{\rm e} x_{\rm e} \\
Y_{20} & = & \omega_{\rm e} y_{\rm e}.
\end{eqnarray}

Up to this point of the analysis, rotational degrees of freedom have not been considered. Quantized rotational energy levels labeled by the angular momentum quantum number $J$ are obtained by adding the spherical-harmonics contribution
\begin{equation}
\frac{J(J+1)}{2mr^2} \nonumber
\end{equation}
to the vibrational Hamiltonian. The diagonalization of the rotationally augmented Hamiltionan yields the energy levels
\begin{equation}
E_{\nu J} = hc \sum_{k,l=0}^{+\infty} Y_{kl}(\nu+1/2)^k(J(J+1))^l,
\end{equation}
which are conventionally rewritten in the Herzberg representation as
\begin{eqnarray}
E_{\nu J} & = & hc\Big[\omega_{\rm e}(\nu+\frac 12) - \omega_{\rm e}x_{\rm e}(\nu+\frac 12)^2 \nonumber \\
&& + \omega_{\rm e}y_{\rm e}(\nu+\frac 12)^3 - \alpha_{\rm e} (\nu+\frac 12) J(J+1) \nonumber \\
&& + B_{\rm e}J(J+1) - D_{\rm e} (J(J+1))^2  + \cdots \Big].
\end{eqnarray}
The coefficient $\alpha_{\rm e}$ that corresponds to the leading anharmonic contribution and couples vibrational degrees of freedom to rotational degrees of freedom is of central significance to the analysis presented in Sec. \ref{Analysis}.

To assess the performance of quantum-mechanical approximations in describing the harmonic and anharmonic vibrational properties of CO in the gas phase, we perform first-principles spectroscopic calculations using the plane-wave {\sc cp} (Car-Parinello) code of the {\sc quantum-espresso} distribution \cite{GiannozziBaroni2009} that optimizes electronic degrees of freedom via damped fictious Newtonian dynamics. \cite{LaasonenPasquarello1993} Our calculations are carried out at three different levels of quantum approximation, namely, uncorrelated Hartree-Fock (HF), the local density approximation (LDA),\cite{PerdewZunger1981} and the semilocal Perdew-Burke-Ernzerhof (PBE)\cite{PerdewBurke1996} approximation with density-gradient corrections. The size of the cubic supercell is set to be of 30 bohr with countercharge corrections to eliminate dipole interactions between artificial periodic images of the CO molecule. \cite{LiDabo2011} In performing gas-phase calculations, we employ norm-conserving pseudopotentials and select the cutoff energy of the wave-function plane-wave decomposition (smooth discretization grid) to be always higher than 30 Ry. Note that we use LDA pseudopotentials in our HF calculations, an approximation that has been justified by directly comparing our HF structural and vibrational results to their tabulated counterparts. \cite{NIST2011} In evaluating spectroscopic parameters, we perform a polynomial regression of order 8. With these calculation parameters, all of the spectroscopic constants are found to be converged within less than a few percents. In particular, we verify that the predicted PBE anharmonic parameter $\alpha_{\rm e}$ is converged within 0.005 cm$^{-1}$, as shown on the two horizontal scales in Fig.~\ref{HerzbergConvergence}.

The results reported in Table \ref{GasPhaseSpectroscopy} confirm the significance of DFT correlation in describing the structural and harmonic properties of CO. As a matter of fact, the predicted LDA bond length $r_{\rm e}= 1.124$ \AA\ and vibrational frequency $\omega_{\rm e}= 2128.59$ cm$^{-1}$ are in closer agreement with experimental data, $r_{\rm e}= 1.128$ \AA\ and $\omega_{\rm e}= 2169.81$ cm$^{-1}$, than their HF counterparts, $r_{\rm e}= 1.087$ \AA\ and $\omega_{\rm e}= 2434.14$ cm$^{-1}$ (in accordance with the known HF propensity to overbind). The same trend is observed for anharmonic properties; including local electron correlation at the LDA level improves significantly the determination of Herzberg constants, reducing the errors by nearly one order of precision in comparison with uncorrelated HF. Semilocal PBE is also more precise than HF. However, PBE is found here to be slightly less accurate than LDA, exhibiting some tendency to underbind CO.

Having confirmed the accuracy of gas-phase predictions in reproducing experimental data, we now proceed to assess the performance of LDA and PBE in describing adsorption geometries and vibrational modes for CO on select close-packed metal surfaces.

\subsection{Adsorbed carbon monoxide}

\label{AdsorbedCarbonMonoxide}

\begin{table*}
\caption{LDA and PBE structural parameters for CO adsorbed on close-packed (111) metal surfaces compared with experiment ($r_{\rm e}$ denotes the equilibrium intramolecular bond length, $r_{\rm m}$ stands for the adsorption distance from the carbon atom to the nearest metal atom, and $\theta$ is the vertical tilt angle of the CO molecule). Unreported computational data correspond to bridge configurations undergoing a transition to the hollow site during structural optimization.} 
\label{AdsorptionProperties}
\begin{tabular*}{0.8\textwidth}{@{\extracolsep{\fill}}cccccccccc}
\hline \hline
 & & \multicolumn{3}{c}{$r_{\rm e}$ (\AA)} & \multicolumn{3}{c}{$r_{\rm m}$ (\AA)} & \multicolumn{2}{c}{$\theta$ ($^\circ$)}                                              \\
Metal & Site & LDA & PBE & Expt. & LDA & PBE & Expt. & LDA & PBE \\
\hline
\\
Ag & atop & 1.140 & 1.148 & --- & 2.053 & 2.173 & --- & 0.8 & 2.6 \\
 & bridge & 1.153 & 1.161 & --- & 2.168 & 2.275 & --- & 0.7 & 2.0 \\
 & hollow & 1.160 & 1.167 & --- & 2.230 & 2.337 & --- & 0.4 & 0.3 \\
\\
Au & atop & 1.138 & 1.148 & --- & 1.960 & 2.044 & --- & 0.6 & 2.4 \\
 & bridge & 1.159 & 1.167 & --- & 2.082 & 2.149 & --- & 1.7 & 2.8 \\
 & hollow & 1.169 & 1.176 & --- & 2.156 & 2.228 & --- & 0.4 & 0.4 \\
\\
Cu & atop & 1.145 & 1.154 & --- & 1.822 & 1.874 & 1.91(1)\footnotemark[1] & 1.3 & 1.2 \\
 & bridge & ---   & 1.170 & --- & ---   & 2.002 & --- & --- & 0.5 \\
 & hollow & 1.168 & 1.178 & --- & 2.000 & 2.062 & --- & 0.7 & 0.6 \\
\\
Pd & atop & 1.143 & 1.152 & --- & 1.849 & 1.891 & --- & 0.4 & 0.2 \\
 & bridge & ---   & ---   & --- & ---   & ---   & --- & --- & --- \\
 & hollow & 1.174 & 1.184 & 1.15(5)\footnotemark[2] & 2.040 & 2.078 & 2.05(4)\footnotemark[2] & 0.2 & 0.1 \\
\\
Pt & atop & 1.144 & 1.154 & 1.15(5)\footnotemark[3] & 1.837 & 1.865 & 1.85(10)\footnotemark[3] & 0.4 & 0.7 \\
 & bridge & 1.167 & 1.177 & 1.15(5)\footnotemark[3] & 2.001 & 2.032 & 2.08(7)\footnotemark[3] & 1.2 & 1.8 \\
 & hollow & 1.178 & 1.188 & --- & 2.084 & 2.120 & --- & 0.4 & 0.4 \\
\\
Rh & atop & 1.148 & 1.159 & 1.20(5)\footnotemark[4]  & 1.825 & 1.855 & 1.87(4)\footnotemark[4] & 0.7 & 0.3 \\
 & bridge & 1.169 & 1.180 & 1.15(10)\footnotemark[5] & 1.988 & 2.022 & 2.03(7)\footnotemark[5] & 1.1 & 1.3 \\
 & hollow & 1.177 & 1.188 & 1.15(10)\footnotemark[6] & 2.059 & 2.097 & 1.99(7)\footnotemark[6] & 0.3 & 0.0 \\
\\
\hline
\hline
\end{tabular*}
\flushleft
\footnotemark[1]{ARPEFS (angle-resolved photoemission extended fine structure technique, Ref.~\onlinecite{MolerKellar1996}).} \\
\footnotemark[2]{LEED (low-energy electron diffraction technique, Ref.~\onlinecite{OhtaniHove1987}).} \\
\footnotemark[3]{LEED (Refs.~\onlinecite{OgletreeVan-Hove1986} and \onlinecite{BlackmanXu1988}).} \\
\footnotemark[4]{LEED (Ref.~\onlinecite{GiererBarbieri1997}).} \\
\footnotemark[5]{LEED (bridge structure obtained in the presence of near-top coadsorbed CO, Ref.~\onlinecite{Van-HoveKoestner1983}).} \\
\footnotemark[6]{LEED (hollow structure obtained in the presence of coadsorbed benzene, Ref.~\onlinecite{OgletreeHove1987}).}
\end{table*}

\begin{table}
\caption{LDA and PBE intramolecular vibrational frequencies $\omega_{\rm e}$ (cm$^{-1}$) for CO adsorbed on close-packed (111) metal surfaces compared with low-coverage experimental data.
\label{AdsorptionFrequencies}}
\begin{tabular*}{\columnwidth}{@{\extracolsep{\fill}}ccccc}
\hline \hline
Metal & Site & LDA & PBE & Expt. \\
\hline
\\
Ag & atop & 2084 & 2026 &  2042(6)\footnotemark[1]                                     \\
             & bridge & 1965             & 1913             & 1970(10)\footnotemark[1] \\
             & hollow & 1919             & 1875             & 1884(21)\footnotemark[1] \\
\\
Au           & atop   & 2107             & 2035             & 2060\footnotemark[2]     \\
             & bridge & 1938             & 1882             & ---                      \\
             & hollow & 1859             & 1817             & ---                      \\
\\
Cu           & atop   & 2073             & 2009             & 2075\footnotemark[3]     \\
             & bridge & ---              & 1873             & ---                      \\
             & hollow & 1887             & 1825             & $<$1840\footnotemark[3]  \\
\\
Pd           & atop   & 2103             & 2033             & $<$2090\footnotemark[4]  \\
             & bridge & ---              & ---              & ---                      \\
             & hollow & 1851             & 1786             & 1840\footnotemark[4]     \\
\\
Pt           & atop   & 2118             & 2048             & 2070\footnotemark[5]     \\
             & bridge & 1917             & 1851             & 1830\footnotemark[5]     \\
             & hollow & 1825             & 1759             & 1760\footnotemark[5]     \\
\\
Rh           & atop   & 2079             &  2004            & 1990\footnotemark[6]     \\
             & bridge & 1899             &   1832           & $<$1870\footnotemark[6]  \\
             & hollow & 1835             &   1771           & ---                      \\
\\
\hline
\hline
\end{tabular*}
\flushleft
\footnotemark[1]{FTIR (Fourier transform infrared spectroscopy, Ref.~\onlinecite{OrozcoPerez1998}).} \\
 \footnotemark[2]{PR-IRAS (polarization-modulation infrared reflection-absorption spectroscopy, Ref.~\onlinecite{Rupprechter2007}).} \\
\footnotemark[3]{IRAS (infrared reflection-absorption spectroscopy, Ref.~\onlinecite{HaydenKretzschmar1985}).} \\
\footnotemark[4]{SFG (sum-frequency generation spectroscopy, Ref.~\onlinecite{RupprechterUnterhalt2002}).} \\
\footnotemark[5]{SFG (Ref.~\onlinecite{LuWhite2004}).} \\
\footnotemark[6]{HREELS (high-resolution electron energy loss spectroscopy, Ref.~\onlinecite{DuboisSomorjai1980}).}
\end{table}

\begin{figure}[ht!]
\includegraphics[width=8cm]{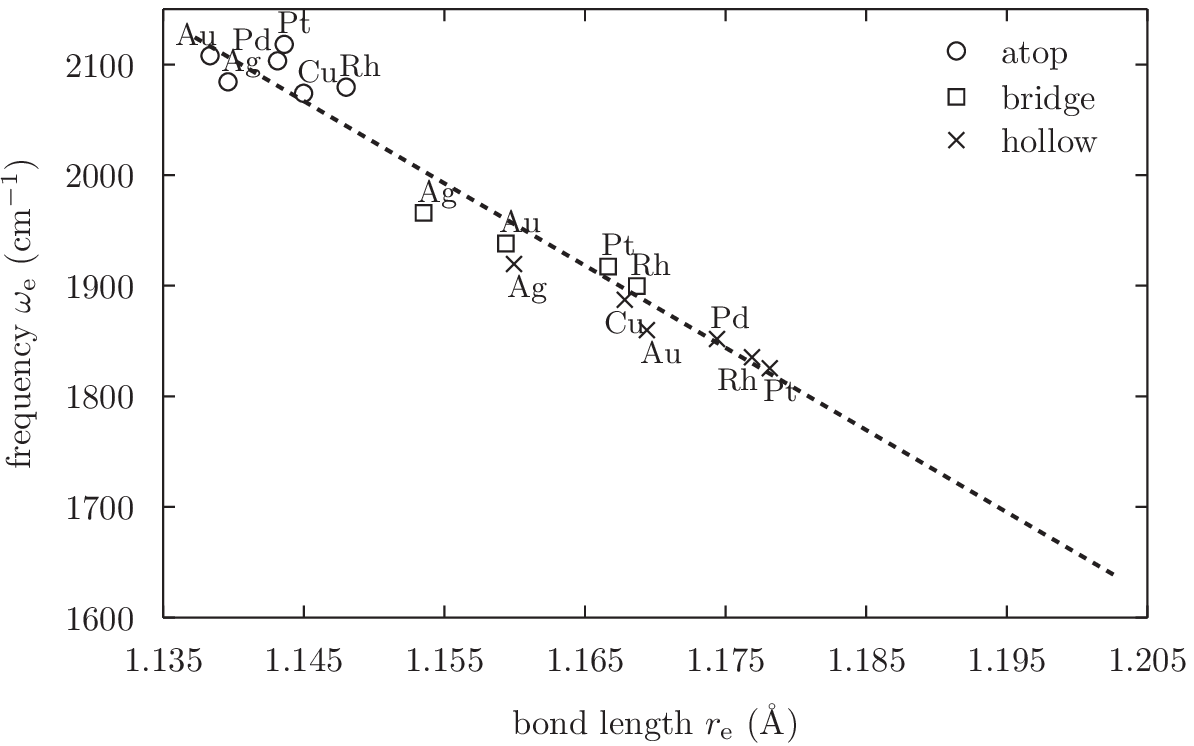}
\caption{
Equilibrium stretching frequency $\omega_{\rm e}$ as a function of the bond length $r_{\rm e}$ within LDA for CO adsorbed at the atop, bridge, and hollow sites on metal surfaces. The linear-regression slope is calculated to be --7659 cm$^{-1}\cdot$\AA$^{-1}$.
\label{CorrelationLDA}}
\end{figure}

\begin{figure}[ht!]
\includegraphics[width=8cm]{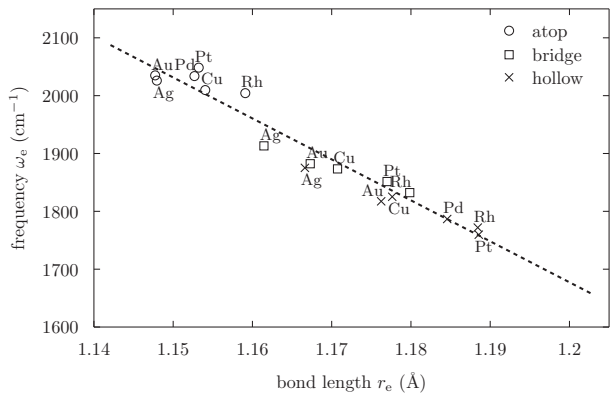}
\caption{
Equilibrium stretching frequency $\omega_{\rm e}$ as a function of the bond length $r_{\rm e}$ within PBE for CO adsorbed at the atop, bridge, and hollow sites on metal surfaces. The linear-regression slope is calculated to be --7151 cm$^{-1}\cdot$\AA$^{-1}$.
\label{CorrelationPBE}}
\end{figure}

Herein, we focus on predicting structural and vibrational properties of CO adsorbed on close-packed surfaces within LDA and PBE using the plane-wave {\sc pw} code of the {\sc quantum-espresso} distribution, which proceeds by iterative diagonalization of the effective Kohn-Sham Hamiltonian. \cite{GiannozziBaroni2009}

At this stage of the study, we emphasize that care must be exerted in applying LDA and PBE to describe CO adsorption. In fact, it is known that local and semilocal DFT approximations occasionally fail in predicting the relative stability of adsorption sites. These qualitative deficiencies take root into spurious electron self-interaction (i.e., the incorrect analytical behavior of the approximate ground-state energy as a function of the electron number, which favors adsorption sites of high coordination). One notorious instance of the LDA and PBE failures is the overstabilization of multifold  adsorption sites (hollow and bridge) relative to terminal adsorption sites (atop) for CO at close-packed platinum surfaces. \cite{FeibelmanHammer2001,KresseGil2003,Abild-PedersenAndersson2007}

In spite of quantitative errors in predicting the adsorption energetics, it has been established from the analysis of the density of states (DOS) and force density of states (FDOS) that semilocal DFT approximations are accurate in determining structural and vibrational properties for adsorbed CO. \cite{DaboWieckowski2007} The calculations presented below corroborate the performance of LDA and PBE in describing structural and vibrational properties.

Our computational benchmark consists of a pool of 6 representative metal elements with face-centered crystal structures, namely, Ag, Au, Cu, Pd, Pt, and Rh. First, we determine lattice parameters and bulk moduli at the different levels of DFT approximation, controlling convergence with respect to the cutoff kinetic energy of the plane-wave expansion of the electronic wave functions (and total electronic density when employing ultrasoft potentials), to the Monkhorst-Pack sampling of the Brillouin zone, and to the temperature of the generalized electronic entropy.\footnote{A wave-function cutoff energy $E_{\rm cut}= 30$ Ry is sufficient to achieve targeted convergence levels for structural, harmonic, and anharmonic properties with the exception of Cu (40 Ry for LDA and 35 Ry for PBE), Pd (40 Ry for PBE), Pt (40 Ry for PBE), and Rh (35 Ry for LDA). In our ultrasoft calculations, the charge-density cutoff energy is set to be $8E_{\rm cut}$. For bulk crystals, the Monkhorst-Pack sampling of the Brillouin zone is set to be 10 $\times$ 10 $\times$ 10 (selecting a Marzari-Vanderbilt generalized-entropy temperature of 0.03 Ry) with the exception of Au (LDA) and Cu (PBE) for which the Brillouin resolution has to be raised to 12 $\times$ 12 $\times$ 12. Based upon calculated lattice parameters, an equivalent surface density of sampling points is applied in adsorption calculations.} 

Predicted crystal properties are reported and compared with experiment in Appendix \ref{CrystalProperties}. Based upon calculated lattice parameters, we construct periodic (111) surface models adopting a fully relaxed orthorhombic $(\sqrt{3}\times 2)$ four-layer slab geometry to determine the vibrational properties of CO adsorbed at the atop, bridge, hollow (face-centered cubic) sites, corresponding to a low coverage of 1/4 ML. We underscore that adopting a model $(\sqrt{3}\times 2)$ geometry is not restrictive for the present analysis as the resulting variations in structural and vibrational predictions are negligible on the scale of the Badger correlation as discussed further below. All of the considered adsorption systems correspond to upright, nondissociative adsorption and are not affected by spin-polarization. In determining and analyzing vibrational frequencies, we employ a frozen-phonon method restricted to vertical C and O atomic displacements. In this frozen-phonon picture, one determines the stretching frequency from the largest eigenvalue of the two-dimensional intramolecular dynamical matrix $D_{ij}=(m_i m_j)^{-\frac 12} \partial_{z_i} \partial_{z_j} V(z_1,z_2)$ where $z_1$ and $z_2$ denote the vertical displacements of the C atom of mass $m_1$ and O atom of mass $m_2$. The restriction to transverse adsorbate coordinates represents a reliable approximation for the intramolecular vibrational mode of light CO on heavy metals, yielding equilibrium stretching frequencies in close agreement with the result of the full DFPT (density-functional perturbation theory)\cite{BaroniGironcoli2001} calculation of the density matrix, with a maximal error of 2\% for CO sitting at the most coordinated (i.e., hollow) site of the lightest (i.e., Cu) substrate. 

In Table \ref{AdsorptionProperties}, we report LDA and PBE structural parameters, namely, the equilibrium bond length $r_{\rm e}$, the distance from the carbon atom to the closest metal atom $r_{\rm m}$, and the angular tilt of the CO molecule $\theta$. Our computational predictions are compared with available experimental data, mainly derived from low-energy electron diffraction (LEED). Of note in Table \ref{AdsorptionProperties} is the lack of experimental structural data for CO adsorbed on noble metals and the difficulty to probe adsorption sites of high coordination without specific coadsorption [as is the case for CO at the hollow site of Rh(111)].\cite{OgletreeHove1987} Notwithstanding experimental uncertainties and variations in adsorption configurations, LDA and PBE predictions always lie within (or very close to) the experimental range. Precisely, LDA tends to underestimate $r_{\rm e}$ by a few hundredths of \AA, whereas PBE exhibits improved predictive performance. Additionally, LDA and PBE calculations confirm the experimentally observed upright adsorption geometry for all of the surfaces considered.

In Table \ref{AdsorptionFrequencies}, local and semilocal intramolecular vibrational frequencies are presented and compared with low-coverage experiments based upon infrared techniques, including infrared reflection-absorption spectroscopy (IRAS), sum-frequency generation (SFG), and high-resolution electron energy loss spectroscopy (HREELS). LDA and PBE calculations predict important vibrational redshifts as a function of the adsorption coordination, as large as of 300 and 345 cm$^{-1}$ for the hollow site of Pt. Instead, the lowest redshifts are observed for CO adsorbed at the terminal site on Pt. From available experimental frequencies, LDA and PBE mean absolute errors are estimated to be of 40 and 30 cm$^{-1}$, respectively. 

With these results in hand, we can graphically confront predicted C--O vibrational frequencies $\omega_{\rm e}$ with equilibrium bond lengths $r_{\rm e}$, thereby recovering the close correlation highlighted in Ref.~\onlinecite{WasileskiWeaver2002}. In quantitative terms, linear regression yields Badger slopes of $-7659$ and $-7151$ cm$^{-1}\cdot$\AA$^{-1}$ within LDA and PBE, respectively (Fig.~\ref{CorrelationLDA} and Fig.~\ref{CorrelationPBE}). It is important to note that our PBE prediction is in very good agreement with the slope of $-7066$ cm$^{-1}$ reported in Ref.~\onlinecite{GajdosEichler2004} using the semilocal PW91 (Perdew-Wang) approximation\cite{PerdewWang1992} with a different adlayer model, confirming the validity of our structural approximation. In the next section, we present a quantitative procedure to rationalize Badger's rule in terms of gas-phase spectroscopic properties. 

\section{Analysis}

\label{Analysis}

\subsection{Analytical procedure}

\label{AnalyticalProcedure}

\begin{figure}[ht!]
\includegraphics[width=8cm]{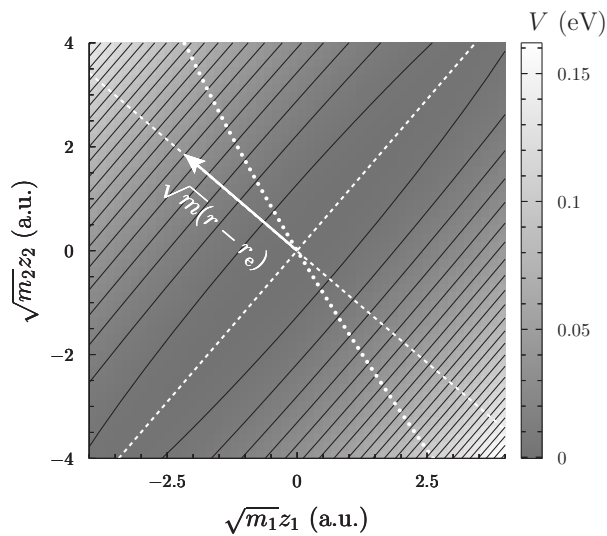}
\caption{
Anharmonic stretching direction (dotted line) in the plane of mass-scaled atomic coordinates $(\sqrt{m_1}z_1,\sqrt{m_2}z_2)$ for CO at the atop adsorption site on Ag. The anharmonic stretching direction connecting energy minima departs from its harmonic counterpart (dashed line with arrow).
\label{CurvedDirection}}
\end{figure}

In a nutshell, the intramolecular analysis developed here consists of deconvoluting the intramolecular potential energy along the curved anharmonic stretching trajectory of CO in adsorption (corresponding to the dotted line in Fig.~\ref{CurvedDirection}).

To determine the anharmonic stretching trajectory, we first calculate the potential energy surface corresponding to constrained vertical displacements of the C and O atoms, the total energy being minimized with respect to all of the other molecular and metal-atom coordinates. For instance, we consider the case of CO adsorbed at the atop site of Ag in Fig.~\ref{CurvedDirection}. The two-dimensional potential energy surface is sampled with 17 points uniformally distributed along the ${\hat {\bf z}}_1$, $\hat {\bf z}_2$, $(\hat {\bf z}_1+\hat {\bf z}_2)/\sqrt 2$ and $( \hat{\bf z}_1-\hat{\bf z}_2)/\sqrt 2$ unit vectors that define the C, O, C--O, and Pt--CO displacement directions. As already mentioned, for light CO adsorbed on heavy metals, restricting the analysis to transverse adsorbate coordinates provides an accurate description of the vibrational behavior of CO with errors that do not exceed 1-2\% in predicting the intramolecular stretching frequency. The stretching trajectory is obtained by connecting the points derived from the minimization of the total energy fixing the intramolecular distance. Minimum-energy points are calculated by simple steepest descent, projecting the energy gradient on the direction defined by the constraint on the bond length. Finally, the total energy is plotted as a function of the constrained bond length, thereby obtaining the intramolecular potential energy curve describing the elongation of adsorbed CO. 

This analytical construction provides a straightforward generalization of harmonic modes to the anharmonic case and offers a direct means to extract the intrinsic vibrational properties of any polyatomic system coupled to a substrate. Note that the curved stretching trajectory reduces to a straight line when the potential energy surface is represented by a parabolic well, in agreement with physical intuition. As a consequence, the curvilinear deviation of the stretching trajectory reflects anharmonic contributions to the intramolecular potential energy of relevance to the present analysis.

\subsection{Intramolecular correlation}

\label{IntramolecularCorrelation}

\begin{figure}[ht!]
\includegraphics[width=7.5cm]{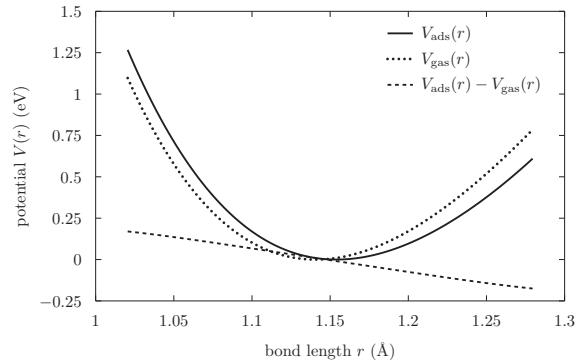}
\caption{Intramolecular potential $V_{\rm ads}(r)$ for CO adsorbed at the atop site of Pt, intramolecular potential $V_{\rm gas}(r)$ in the gas phase, and their difference $V_{\rm ads}(r)-V_{\rm gas}(r)$.
\label{AdsorptionContribution}}
\end{figure}

\begin{figure}[ht!]
\includegraphics[width=8cm]{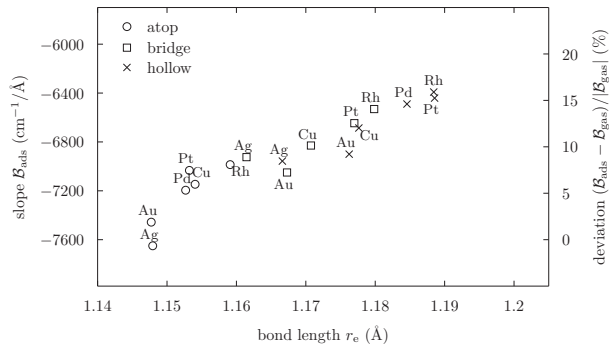}
\caption{
Badger slope ${\cal B}_{\rm ads}$ (left scale) and percent deviation of the Badger slope  (right scale) relative to its gas-phase counterpart as a function of the bond length $r_{\rm e}$ for CO adsorbed at the atop, bridge, and hollow sites of metal surfaces.
\label{SlopePBE}}
\end{figure}

Using the procedure outlined above, we now examine the contribution from the adsorption environment to the intramolecular potential. For this analysis, we focus on semilocal PBE, which has been shown above to be more accurate than local LDA in predicting the vibrational properties of adsorbed CO.

In Fig.~\ref{AdsorptionContribution}, the intramolecular potential $V_{\rm ads}(r)$ for CO at the atop site of Pt is compared to the intramolecular potential $V_{\rm gas}(r)$ for CO in the gas phase. Evaluating the difference between the two potentials, we obtain the adsorption contribution to the intramolecular potential, which is close to linear on a large range spanning all of the bond lengths $r_{\rm e}$ reported in Table~\ref{AdsorptionProperties}. This provides computational evidence for the preservation of the anharmonicity of the intramolecular potential despite strong electronic hybridization.

To extend this observation, we now consider the other adsorption systems described in Sec. \ref{AdsorbedCarbonMonoxide}. In Fig.~\ref{SlopePBE} (left scale), we calculate the Badger slope ${\cal B}_{\rm ads}$ at the minimum of the intramolecular potential:
\begin{equation}
{\cal B}_{\rm ads} \equiv \left. \frac{{\rm d} \omega_{\rm ads}}{{\rm d} r} \right|_{r=r_{\rm e}},
\end{equation}
where $\omega_{\rm ads}(r)$ is related to the second derivative of the potential through
\begin{equation}
\frac{{\rm d}^2 V_{\rm ads}}{{\rm d} r^2} = mc^2(2\pi \omega_{\rm ads})^2.
\end{equation}
We first observe that the influence of the substrate drops as the activity of the metal and coordination of the adsorption site decrease; the contribution from the substrate is found to be maximal at the hollow site of Rh and Pt while is minimal at the atop site of Ag and Au. Despite these fluctuations, the magnitude of the slope ${\cal B}_{\rm ads}$ remains within 6400-7600 cm$^{-1}\cdot$\AA$^{-1}$, explaining the correlation between $\omega_{\rm e}$ and $r_{\rm e}$ depicted in Fig.~\ref{CorrelationPBE}. 

To gain further quantitative insight, we compare calculated Badger slopes ${\cal B}_{\rm ads}$ to the Badger slope of the isolated molecule
\begin{equation}
{\cal B}_{\rm gas} \equiv \left. \frac{{\rm d} \omega_{\rm gas}}{{\rm d} r} \right|_{r=r_{\rm e}},
\end{equation}
where $\omega_{\rm gas}(r)$ is related to the second derivative of the potential in the gas phase.
The slope ${\cal B}_{\rm gas}$ can be calculated from the potential energy $V_{\rm gas}(r)$, yielding a correlation coefficient of $-7601$ cm$^{-1}\cdot$\AA$^{-1}$. Alternatively, ${\cal B}_{\rm gas}$ can be evaluated by exploiting the Dunham expansion [Eq.~(\ref{DEP})]. As a matter of fact, making use of the relation
\begin{equation}
\alpha_{\rm e}=- \frac{6B_{\rm e}^2}{\omega_{\rm e}} \left(1+a_1\right) + \cdots
\end{equation}
that is valid to leading order in the adimensional number $B^2_{\rm e}/\omega^2_{\rm e}$ [see Eq.~(15) in Ref.~\onlinecite{Dunham1932}], one arrives at the analytical expression
\begin{equation}
{\cal B}_{\rm gas} = - \frac{3 \omega_{\rm e}}{2r_{\rm e}} \left(1+\frac{\omega_{\rm e} \alpha_{\rm e}}{6 B_{\rm e}^2}+\cdots\right). 
\label{ICS}
\end{equation}
Substituting the terms in Eq.~(\ref{ICS}) for the calculated values reported in Table~\ref{HerzbergConvergence}, we obtain a slope of $-7600$ cm$^{-1}\cdot$\AA$^{-1}$, which is in very good agreement with the value of $-7601$ cm$^{-1}\cdot$\AA$^{-1}$ determined above. By comparing this slope to that of adsorbed CO [Fig.~\ref{SlopePBE} (right scale)], we conclude that the magnitude of the surface contribution does not exceed 16\%  and is even lower than 5\% for terminal sites on noble metals. 

The central conclusion of this analysis is that the hybridization of CO molecular orbitals with metal bands causes a maximal departure of 1200 cm$^{-1}\cdot$\AA$^{-1}$ from the gas-phase Badger correlation. This trend provides quantitative justification of Badger's rule and allows us to express the slope ${\cal B}_{\rm ads}$ for CO in adsorption in terms of the anharmonic constants of the isolated molecule. In the next section, we re-examine this conclusion taking into account the influence of electrical conditions and we discuss the validity of conventional adsorption models in describing electrified surfaces.

\section{Discussion}

\label{Discussion}

\begin{figure}[ht!]
\includegraphics[width=8cm]{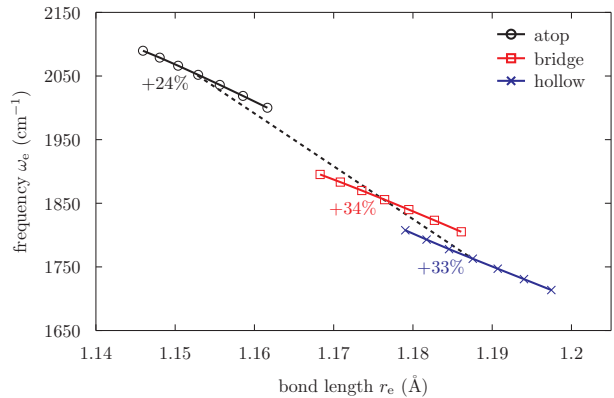}
\caption{
Intramolecular frequency $\omega_{\rm e}$ as a function of the bond length $r_{\rm e}$ for CO adsorbed at the atop, bridge, and hollow sites of Pt considering a varying surface charge. Departure from Badger's rule is expressed in terms of percent deviations of the charge-dependent slope ${\cal B}_{\rm elec}$ relative to the gas-phase slope ${\cal B}_{\rm gas}$. The dashed line serves as a guide for the eye.
\label{CorrelationCharge}}
\end{figure}

The intramolecular analysis has demonstrated the possibility of embedding the complexity of the adsorption environment into a simple linear constraint that preserves Badger's rule. Based upon this observation, it is tempting to infer that Badger's rule is also conserved under electrical conditions. Here, we confront commonly used adsorption models, namely, finite-cluster and periodic-slab simulations under electrical conditions to discuss the validity of this conjecture. 

We first concentrate on finite-cluster models, which have been employed in simulating adsorption in the low-coverage limit. \cite{WasileskiKoper2001, WasileskiWeaver2001, WasileskiKoper2002, WasileskiWeaver2002} To analyze the influence of electrical conditions within finite-cluster models, we exploit data from Ref.~\onlinecite{WasileskiWeaver2001}, which have been computed at the semilocal generalized-gradient level using a finite Pt$_{13}$ cluster consisting of 2 hexagonal layers of 7 and 6 atoms; such a Pt$_{13}$-cluster structure allows us to study atop adsorption on the 7-atom facet and hollow adsorption on the 6-atom facet. In these calculations, the influence of the electric field is accounted for via a transverse linear potential added to the effective electronic Hamiltonian and interatomic potential. Based upon extracted data (cf. Fig.~2 and 3 of Ref.~\onlinecite{WasileskiWeaver2001}), we calculate the field-dependent slope
\begin{equation}
{\cal B}_{\rm elec} \equiv \left. \frac{{\rm d} \omega_{\rm e}}{{\rm d} {\sf E}} \right|_{{\sf E}=0} \left( \left. \frac{{\rm d} r_{\rm e}}{{\rm d} {\sf E}} \right|_{{\sf E}=0} \right)^{-1}
\end{equation}
to be 6790 cm$^{-1}\cdot$\AA$^{-1}$ at the atop site and 6180 cm$^{-1}\cdot$\AA$^{-1}$ at the hollow site. The calculated slopes are in very good agreement with the slopes ${\cal B}_{\rm ads}$ reported in Sec.~\ref{IntramolecularCorrelation} (Fig.~\ref{SlopePBE}). In quantitative terms, ${\cal B}_{\rm elec}$ deviate by less than 5\% from ${\cal B}_{\rm ads}$ for both atop and hollow adsorptions. These results provide evidence that applying a uniform electric field within finite-cluster simulations conserves Badger's rule.  

Having verified that Badger's rule is preserved within finite-cluster models, we now turn to periodic-slab calculations. \cite{LozovoiAlavi2003, OtaniSugino2006, TaylorWasileski2006, KarlbergRossmeisl2007, LozovoiAlavi2007, JinnouchiAnderson2008, DaboKozinsky2008, DaboKozinsky2011, HamadaMorikawa2008, DeshlahraWolf2009,  MamatkulovFilhol2011} In periodic-slab calculations, electrical conditions can be imposed by applying an external electric field $\sf E$ similarly to finite-cluster calculations (the electric-field method) or, alternatively, by varying the surface charge $\sigma$ through adjusting the number of electrons in the system (the surface-charge method).\cite{DeshlahraWolf2009} For periodic slabs of sufficient thickness, the two approaches can be shown to be equivalent. \cite{DeshlahraWolf2009} Here, we adopt the surface-charge approach with density-countercharge techniques\cite{DaboKozinsky2008} to impose two-dimensional boundary conditions compatible with the slab periodicity. We consider slabs comprising 3 metal layers and verify that increasing the thickness to 4 layers affects Badger-slope predictions by less than 150 cm$^{-1}\cdot$\AA$^{-1}$. The magnitude of the surface charge is raised up to $0.1$ C$\cdot$m$^{-2}$. \footnote{1 C$\cdot$m$^{-2}$ of surface charge corresponds to 11.29 V$\cdot$A$^{-1}$ of vacuum electric field.} As depicted in Fig.~\ref{CorrelationCharge}, we calculate charge-dependent slopes ${\cal B}_{\rm elec}$ by simultaneously monitoring the vibrational frequency $\omega_{\rm e}$ and the intramolecular distance $r_{\rm e}$. We thus obtain charge-dependent slopes of $-5700$ cm$^{-1}\cdot$\AA$^{-1}$ at the atop site, $-5070$ cm$^{-1}\cdot$\AA$^{-1}$ at the bridge site, and $-5010$ cm$^{-1}\cdot$\AA$^{-1}$ at the hollow site. Consequently, calculated slopes are found to deviate significantly from their gas-phase counterparts, demonstrating that Badger's rule does not hold under electrical conditions using periodic-slab models at variance with finite-cluster simulations. 

To explain these trends, we resort to effective electrical variables, which provide a conceptually useful, direct mapping for describing the influence of electrical conditions on physical systems. Explicitly, we first write the free energy of the adsorption system as
\begin{equation}
U(r,{\sf E})=V(r)+S {\sf P}(r,{\sf E}) {\sf E}
\end{equation} 
in terms of the polarization of the adlayer per unit surface area ${\sf P} = {\partial U}/{\partial (S{\sf E})}$. Variations with respect to the intramolecular distance $r$ yields an implicit electric-field dependence for the equilibrium bond length of the form ${{\rm d} V}/{{\rm d} r}|_{r=r_{\rm e}({\sf E})} + S\sigma^\star(r_{\rm e}({\sf E}),{\sf E}) {\sf E} = 0$ where
\begin{equation}
\sigma^\star = \frac{\partial {\sf P}}{\partial r}
\end{equation} 
is known as the {\it Born effective charge} per surface area and is commonly used in describing the coupling between structural properties and an external electric field. \cite{BaroniGironcoli2001} Making use of this equilibrium relation, we derive the expression of the field-dependent Badger slope
\begin{equation}
{\cal B}_{{\rm elec}} = {\cal B}_{\rm ads} -\frac  {\omega_{\rm e}}2 \left. \frac{\partial \ln |\sigma^\star|}{\partial r}  \right|_{r=r_{\rm e}, {\sf E}=0},
\label{BFD}
\end{equation}
which shows that ${\cal B}_{{\rm elec}}$ departs from ${\cal B}_{{\rm ads}}$ by a term that is related to the logarithmic derivative of the effective charge $\sigma^\star(r,{\sf E})$ as a function the bond length.

As a result, the clear accordance between ${\cal B}_{{\rm elec}}$ and ${\cal B}_{{\rm ads}}$ within cluster models indicates that the effective charge of the adlayer remains constant upon stretching the CO adsorbate. In other words, within finite-cluster descriptions, the dipole ${\sf P}(r,{\sf E})$ depends linearly on the bond length $r$ due to the absence of charge redistribution at the surface of the cluster. 

In notable contrast to finite clusters, a nonlinear coupling mechanism takes place on periodic slabs; substituting our PBE predictions into Eq.~(\ref{BFD}), we can assess the coefficient of the leading term $\gamma$ that describes the redistribution of the adlayer charge upon elongating the CO molecule, 
\begin{equation}
\sigma^\star = \sigma^\star_{\rm e} \left(1+\gamma \frac{r-r_{\rm e}}{r_{\rm e}}+\cdots \right).
\end{equation}
where $\sigma^\star_{\rm e} = \sigma^\star(r_{\rm e},{\sf E}=0)$. The adimensional coefficient $\gamma$ is calculated to lie between $-2$ at the hollow site and $-1.5$ at the atop site, reflecting the nonlinear dependence of the surface dipole as a function of the bond length. This is at variance with finite-cluster models that fail to capture the decay of the effective charge. In conclusion, finite-cluster models are not apt at describing the structural and vibrational properties of CO adsorbates under varying electric field, strongly suggesting that they should be employed with caution in simulating electrical conditions at charged electrodes in contrast to periodic-surface models that correctly describe the decaying effective surface charge of the adlayer.

\section{Conclusion}

In summary, Badger's rule, which correlates linearly the equilibrium stretching frequency $\omega_{\rm e}$ to the bond length $r_{\rm e}$ of CO under varying adsorption conditions arises from the conservation of intramolecular anharmonicity despite strong chemisorption. The intramolecular analysis has shown that the hybridization of CO molecular orbitals with the metal electronic states causes moderate deviations from the gas-phase Badger correlation for all of the metal electrodes considered; in particular, Badger deviations become negligible for adsorption at the terminal site of noble metals. 

Applying a uniform electric field within finite-cluster models has been shown to not affect Badger's rule significantly, whereas the anharmonic contribution from an electrical conditions within periodic-slab models has been found to alter the intramolecular correlation, ultimately leading to the breakdown of Badger's rule. Analyzing the influence of electrical conditions in terms of effective electrical variables reveals that finite-cluster electric-field models are not apt at capturing variations in the effective surface charge upon elongating the CO adsorbate, explaining their failure in recovering the influence of electrical conditions on the intramolecular bond. 

This first-principles study  underscores the relevance of the redistributed surface charge in understanding the coupling of the adsorbate vibrational modes with the surface electric field and points out limitations of finite-cluster models in simulating adsorption under electrical conditions. These results motivate further theoretical and computational efforts towards more realistic simulations of electrode interfaces --- ideally taking into account the self-consistent polarization and ionic response of the embedding electrolyte. \cite{AndreussiDabo2012} Such computational approaches would find fruitful applications in studying molecular electrochemical phenomena relevant to catalysis and energy conversion. 

\acknowledgments

The author is grateful to N. Bonnet, N. Marzari,  J.-S. Filhol, and A. Kachmar for valuable discussions. The author acknowledges support by Grant ANR 06-CIS6-014 of the French National Agency of Research.

\appendix

\section{Crystal data}

\label{CrystalProperties}

For completeness, we report calculated lattice parameters and bulk moduli together with experimental data in Table \ref{CrystalData}. We observe that lattice parameters are underestimated by LDA with an absolute error of approximately 2\% and overestimated by PBE with an error that does not exceed 3\%. Despite uncertainties in experimental bulk moduli, errors in predicting bulk moduli are admittedly nonnegligible, reaching 35\% for Ag within LDA and $30$\% for Au within PBE. Although inaccuracies in predicting the elastic properties of the substrate may affect the description of vibrational modes involving displacements of metal atoms, the C–O stretching mode is weakly coupled to the substrate and is thus moderately affected by such errors, as shown in Sec.~\ref{AdsorbedCarbonMonoxide}.

\begin{table}
\caption{LDA and PBE lattice parameters and bulk moduli for Ag, Au, Cu, Pd, Pt, and Rh in the face-centered-cubic crystal structure compared with experiment (Ref.~\onlinecite{CRC2011}).
\label{CrystalData}}
\begin{tabular*}{\columnwidth}{@{\extracolsep{\fill}}ccccccc}
\hline \hline
       &             & \multicolumn{2}{c}{DFT}             & \multicolumn{2}{c}{Expt.}                                            \\
Metal  &             & $a$ (\AA)   & $B$ (GPa)             & $a$ (\AA) & $B$ (GPa)                                                \\
\hline
\\
Ag     & LDA         & 4.019       & 135.9                 &  \multirow{2}{*}{4.086}      & \multirow{2}{*}{100}                  \\
       & PBE         & 4.163       & 90.7                                                                                         \\
\\
Au     & LDA         & 4.048       & 191.1                 & \multirow{2}{*}{4.078}      & \multirow{2}{*}{200$\pm$20}            \\
       & PBE         & 4.173       & 140.0                                                                                        \\
\\
Cu     & LDA         & 3.552       & 171.6                 & \multirow{2}{*}{3.615}      & \multirow{2}{*}{140}                   \\
       & PBE         & 3.674       & 127.7                                                                                        \\
\\
Pd     & LDA         & 3.864       & 224.7                 & \multirow{2}{*}{3.890}      & \multirow{2}{*}{180}                   \\
       & PBE         & 3.969       & 169.7                                                                                        \\
\\
Pt     & LDA         & 3.921       & 298.6                 & \multirow{2}{*}{3.924}      & \multirow{2}{*}{230}                   \\
       & PBE         & 3.995       & 246.8                                                                                        \\
\\
Rh     & LDA         & 3.784       & 310.5                 & \multirow{2}{*}{3.803}      & \multirow{2}{*}{327.5$\pm$52.5}        \\
       & PBE         & 3.859       & 259.3                                                                                        \\
\\
\hline
\hline
\end{tabular*}
\end{table}

\end{document}